\documentstyle[12pt,epsf]{article}

\begin{document}
\voffset= -2cm
\setlength{\unitlength}{1mm}
\textwidth 15.0 true cm
\textheight 22.0 true cm
\headheight 0 cm
\headsep 0 cm
\topmargin 0.4 true in
\oddsidemargin 0.25 true in

\newcommand{\beq}{\begin{equation}}

\newcommand{\eeq}{\end{equation}}

\begin{flushright}

ITEP/TH-27/00\\


hepth/xxyyzzz

\end{flushright}

\vspace{0.5cm}

\begin{center}

{\Large\bf Big Bang in $AdS_{5}$
with external field and flat 4d Universe}\\

\end{center}

\bigskip

\begin{center}

{\bf A.S.Gorsky and K.G.Selivanov}

\bigskip

{ ITEP, Moscow, 117259, B.Cheryomushkinskaya 25}

\end{center}

\bigskip

\begin{abstract}

We describe spontaneous creation of the Brane World in $AdS_{5}$
with external field. The resulting Brane World consists of
a flat 4d spatially finite expanding Universe and curved expanding
"regulator" branes. No negative tension branes are involved.

\end{abstract}

1. Recently a lot of attention has been paid to the possibility
to have large \cite{dvali} and even infinite
\cite{rs2} extra dimensions instead of traditional
Kaluza-Klein picture where extra dimensions are extremely small
and thus invisible at moderate energy scales.

The idea of those considerations is that matter \cite{rs1},
gauge fields \cite{pw} and gravity \cite{rs2} can be a sort of
localized on branes (in the present context branes can be viewed
on as fundamental branes or effective domain walls) so that, say,
Newton law is the same as in flat
4d world with small corrections beyond the reach of the
current experimental physics.

A particular construction in \cite{rs2} included
two slices of $AdS_{5}$ matched with each other along
two 3-branes (i.e. branes with 4d world-sheet).
The branes were
localized on a circle in 5th dimension, radius of the circle
being arbitrarily large.
The slices of $AdS_{5}$
were taken in such a way that the 3-branes  were flat
in either of them.
Main feature of this model is that when parameters of the model
(cosmological constant in $AdS_{5}$ and tension of the brane) are tuned
to guaranty the existence of the
stable brane located along the flat section of AdS, one automatically
obtains zero cosmological constant in 4d world on the brane.
A disadvantage of the model (as well as of its modifications,
see e.g. modifications with three branes, two of which have negative
tensions \cite{grs} and some others
\cite{more}) is the necessity to have
auxiliary brane(s) with
negative tension. The negative tension branes (regulator branes, R-branes)
have been realized as a
crucial problem of the construction (see, e.g. \cite{witten}),
and  attempts have been taken to overcome it \cite{dvali2} or
even to abandon the Brane World in favor of the other ideas
about vanishing cosmological constant,
e.g. the one of \cite{bt} which was recently revisited in
\cite{pol}.

In a recent paper \cite{gs} the present authors have suggested a tunneling
mechanism resulting
in a brane configuration looking very similar to the configurations
in \cite{rs2}, \cite{grs}.
The R-branes in the configuration produced are not parallel
to the physical brane (RS-brane) and rapidly
expand. RS-brane is spatially finite, the size being of the
same order as the distance to R-brane. What is important that
none of the branes has negative tension. A peculiar feature of the
configuration is that RS-brane is not located along the flat section
of AdS, hence the 4d cosmological constant is not zero, and even
4d localization of gravity on RS-brane, perhaps, requires additional
justification.

In the present letter we describe a tunneling into the Brane World
such that the resulting RS-brane is a peace of the same
flat section  of $AdS_{5}$
as in \cite{rs2}.
Thus our 4d Universe living on RS-brane is flat and spatially finite
and we can take for granted from
\cite{rs2} the 4d localization of gravity on the RS-brane and
the validity of 4d Newton law far enough in the future.
RS-brane is restricted by junction manifold where RS-brane meets with
rapidly expanding R-branes.

The setup of our consideration is as follows. Originally,
one has $AdS_{5}$ with
homogeneous 4-form field $B$ (so that its curvature
$H=dB$ is proportional to the volume form with a constant coefficient).
Brane production in
that context has been studied in \cite{mms}
(see also \cite{Cvetic}). The study essentially
reduces to the study of minimal charged surfaces in AdS with
the homogeneous field. These surfaces were (up to minor
subtleties) described in \cite{mms}. They are classified into three
classes: undercharged ones (saturating a sort of BPS inequality
between charge and tension of the brane),
overcharged ones (breaking the BPS inequality) and BPS ones (see fig.1-4).
Actually, only overcharged ones were in \cite{mms} given a tunneling
interpretation, the others have infinite volume. Our innovation compared
to \cite{mms} is to include junctions of those surfaces 
(junctions in
in tunneling  were first encountered in \cite{sv}, in \cite{gs1}
junctions were extensively used in description of the induced brane pair 
production). Our Big Bang bounce is glued out of three peaces of branes
- of BPS one, of undercharged one and of overcharged one (see fig.5). The BPS
brane plays the role of RS-brane, the others are R-branes. 
Notice that the overcharged branes inherit the
metric of the sphere, the undercharged branes
- the one of AdS, and the BPS brane - the flat one, we shall
discuss this point in more details later.

The paper is organized as follows. First we describe the charged
minimal surfaces in AdS with external field. Then we construct
the Big Bang bounce describing tunneling into the Brane World.
Finally we present our conclusions. Throughout the paper we use
the test brane approximation - the same approximation as in
\cite{mms} - that is we neglect the brane back reaction on the gravity
and field. The construction is straightforwardly generalizable beyond
this approximation along the lines of \cite{gs}.

2. In this section we describe, essentially
following \cite{mms}, "Euclidean"
minimal charged surfaces, that is,  solutions of test brane worldsheet
equations relevant for tunneling (the ideas of that kind of consideration
of tunneling were developed in \cite{kov}).
Effective action of the test brane can be taken in the following
form:

\beq
\label{action1}
S=T\oint\sqrt g + Q \oint B
\eeq
where $T$ stands for tension of the brane, $Q$ stands for its charge,
$g$ is induced metric on the brane and $B$ is a $(d-1)$-form field.
The integrals are over brane worldsheet.
If one takes the metric of AdS as in \cite{mms},
\beq
\label{metric}
ds^2 =  R_{ads}^2 \left(\cosh^2\!\rho \, d\tau^2 + d\rho^2 + \sinh^2\!{\rho} \,
d\Omega_{d-2}^2 \right),
\eeq
where $R_{ads}$ is the anti-de Sitter radius, and assumes that
the curvature form $H=dB$ of the $B$-field is proportional to the volume form
with a constant coefficient (flux density), and also assumes the spherical symmetry
of the brane worldsheet, one reduces Eq.(\ref{action1})
as follows \cite{mms}:
\beq
\label{action2}
S = TR_{ads}^{d-1} \Omega_{d-2}  \int d\tau \left[
\sinh^{d-2}\rho  \sqrt{ \cosh^2\rho + \left({d \rho
\over d \tau}\right)^2 } - q \sinh^{d-1}\rho  \right]
\eeq
where
$\Omega_{d-2}={2 \pi^{{(d-1)\over 2}} \over \Gamma({d-1 \over 2})}$
stands for the  volume of a unit d-2 sphere
and  $q$ is a constant made out of the flux density, brane charge
and the brane tension. $q=1$ is identified in \cite{mms} as the BPS case.
The branes with $q<1$ will be referred to as undercharged ones and
those with $q>1$ - as overcharged.

Before describing them we would like
to relate the metric Eq.(\ref{metric}) used in \cite{mms} to more
canonical ones. Upon the change of variables,
\beq
\label{variables2}
\tanh \tau = \tan \theta
\eeq
the metric  Eq.(\ref{metric}) is put to the form
\beq
\label{metric2}
ds^2 =  R_{ads}^2 \frac{d\tau^2 + d\theta^{2} + \sin^{2}\theta \,
d\Omega_{d-2}^2 }{\cos^{2}\theta}
\eeq
Finally, upon the change of variables
\beq
\label{variables3}
z=e^{\tau} \cos\theta,\;r=e^{\tau} \sin\theta
\eeq
one obtains the following canonical form of AdS metric
\beq
\label{metric3}
ds^2 =  R_{ads}^2 \frac{dz^2 + dr^{2} + r^{2}\,
d\Omega_{d-2}^2 }{z^{2}}.
\eeq

Let us first consider minimal surfaces for the case with $q<1$
which look as follows
(see also fig.1):
\beq
\label{under}
\cosh\rho = { \sinh \tau_{m} \over \sinh (\tau +a) }
\eeq
where $\tanh \tau_{m}=q$.

\begin{figure}
\hspace*{1.3cm}
\epsfxsize=11cm
\epsfbox{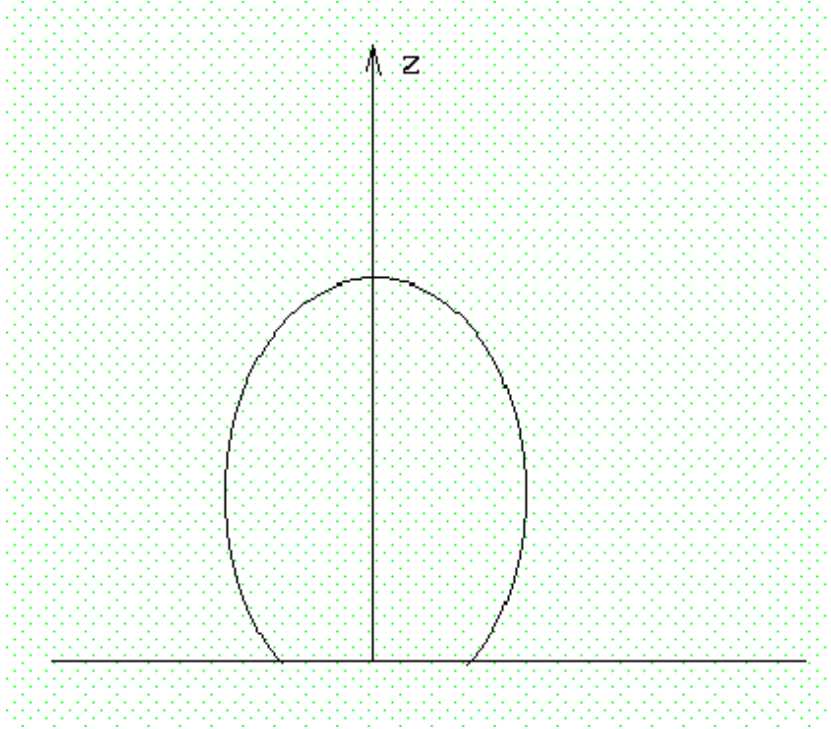}
\caption{Undercharged brane.}
\end{figure}

Notice that we have added
a modulus $a$ compared to Eq.(4.8) in \cite{mms}.
In the context of \cite{mms} this modulus represents the time
translation symmetry and it is more or less irrelevant. In terms
of metric Eq.(\ref{metric3}) this modulus rather defines a scale.
The moduli are necessary to describe the correct limit to the BPS
case ($a$ is sent to infinity when $q \rightarrow 1$) and
they are also necessary to construct the Big Bang bounce with
junctions which we are describing below.
Notice also that in our construction below it is
one of the coordinates orthogonal to $z$ which is
interpreted as Euclidean time, unlike \cite{mms}, where the role
of Euclidean time is played by $\tau$.

In coordinates Eq(\ref{variables2}) the undercharged surfaces
take the form
\beq
\label{under2}
\cos\theta=\frac{\sinh(\tau + a)}{\sinh\tau_{m}}.
\eeq
As we pointed in the Introduction, restriction of the $AdS_{5}$
metric Eq.(\ref{metric2}) onto the undercharged surfaces Eq.(\ref{over2})
gives the metric of $AdS_{4}$.

Let us now turn to the case with $q>1$.
The relevant charged minimal surfaces look as follows
(see also fig.2):
\beq
\label{over}
\cosh\rho = { \cosh \rho_{m}\over \cosh (\tau +b) }
\eeq

\begin{figure}
\hspace*{1.3cm}
\epsfxsize=11cm
\epsfbox{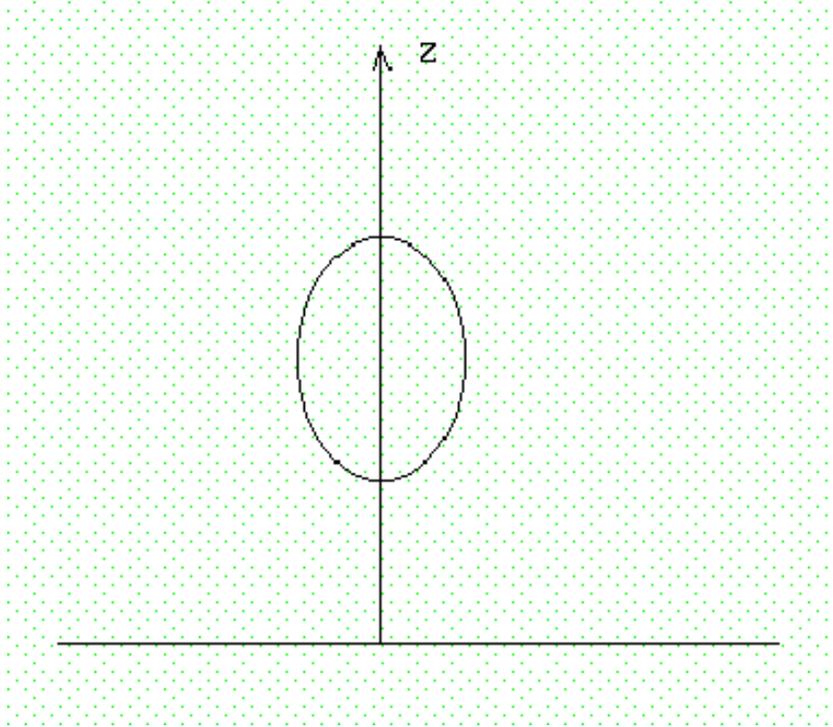}
\caption{Overcharged brane.}
\end{figure}

where  $\tanh \rho_{m} = 1/q$. We have again added
a modulus $b$ compared to Eq.(4.5) in \cite{mms}.
In coordinates Eq(\ref{variables2}) the overcharged surfaces
take the form
\beq
\label{over2}
\cos\theta=\frac{\cosh(\tau + b)}{\cosh\rho_{m}}.
\eeq
As we pointed in the Introduction, restriction of the $AdS_{5}$
metric Eq.(\ref{metric2}) onto the overcharged surfaces Eq.(\ref{over2})
gives the metric of the sphere $S_{4}$.

The BPS case (q=1) can be obtained as a limit from either of the cases
above. Corresponding surfaces look as follows (see also fig.3):

\begin{figure}
\hspace*{2cm}
\epsfxsize=9cm
\epsfbox{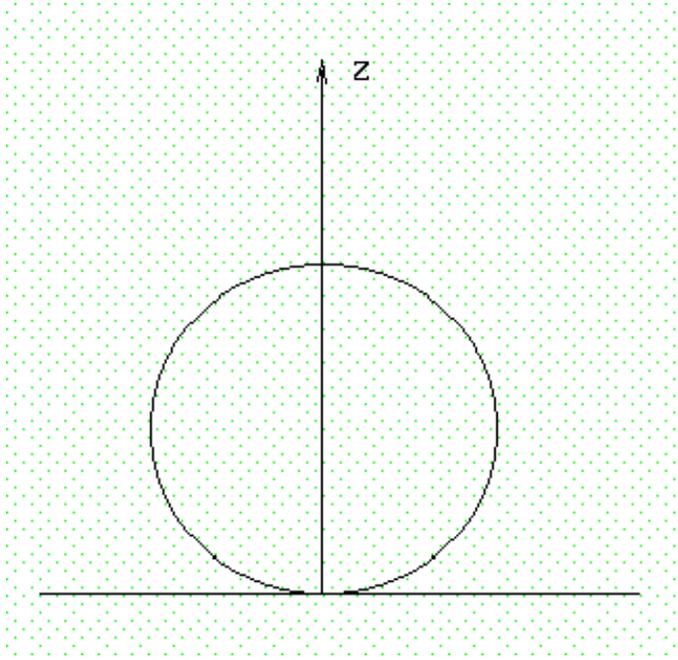}
\caption{BPS brane.}
\end{figure}

\begin{figure}
\hspace*{2cm}
\epsfxsize=9cm
\epsfbox{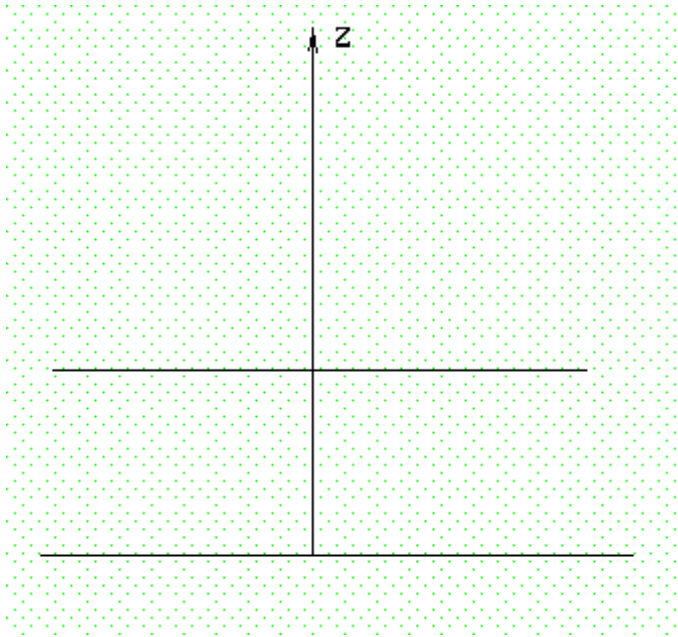}
\caption{BPS brane. It is obtained from the previous case by inversion.}
\end{figure}

\beq
\label{bps}
\cos\theta=\frac{1}{z_{0}}e^{\tau}
\eeq
where $z_{0}$ is a constant.
Upon the inversion transformation one obtains
\beq
\label{bps2}
e^{\tau} \cos\theta=z_{0}
\eeq
In terms of Eq.(\ref{metric3}) these are the surfaces $z=z_{0}$ (fig.4).
Apparently, restriction of the AdS metric onto these surfaces
gives flat Euclidean metric.

Only overcharged surfaces admit a tunneling
interpretation \cite{mms},
since undercharged and BPS ones reach the boundary of AdS
space and thus have infinite volume and infinite effective action.\\

3. We shall now construct the bounce which describes tunneling into the
Brane World. It is glued out of three pieces - a piece of BPS brane
located along $z=z_{0}$ section, Eq.(\ref{bps2}), and playing a role
of RS-brane in the
Brane World, a piece of undercharged brane, Eq.(\ref{under2}),
located above the
BPS brane (in the fixed coordinates of the type of Eq.(\ref{metric3})
and playing a role of one R-brane, and a piece of overcharged brane,
Eq.(\ref{over2}),
located below the BPS brane and playing the role of the other R-brane
(see fig.5). All three pieces are glued along the junction manifold.
The usual junction conditions are the charge conservation and
the tension forces balance at the junction \cite{john} .
We shall assume that there is no junction energy contribution
to the effective action
(such contributions were studied in \cite{junction}
in the context of specific central charges), 
though this assumption is not crucial for the existence of solution.

\begin{figure}
\hspace*{1.3cm}
\epsfxsize=11cm
\epsfbox{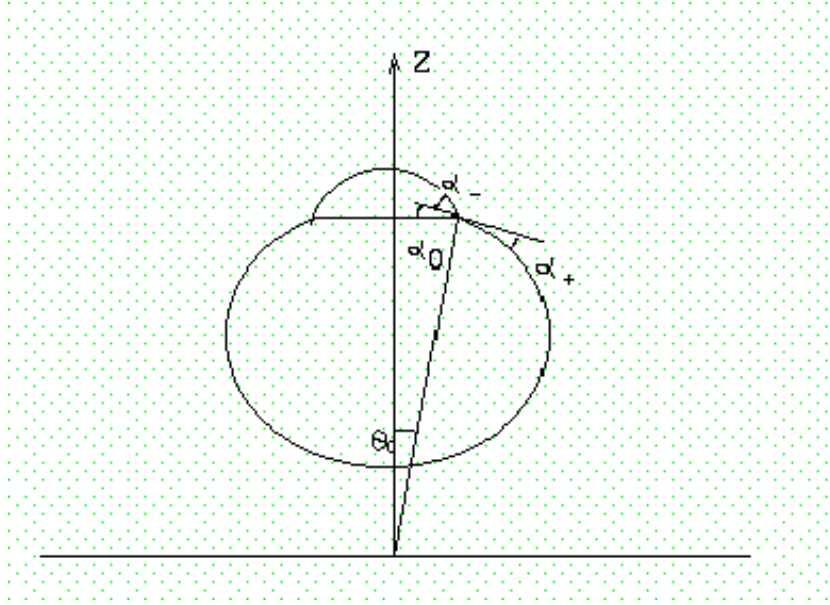}
\caption{Big Bang bounce, or tunneling from AdS to the Brane World.}
\end{figure}

Apparently, the configuration sketched above has a finite effective action 
since none of the constituting pieces reaches the AdS boundary ,
hence the tunneling goes with a finite probability which
can be easily computed. Notice that analogously to constructions
in \cite{mms} one needs a brane breaking BPS inequality in order
to have a finite probability of tunneling. An example of this type of brane
in string theory was given in \cite{mms}.

Junction manifold is of the type  $\theta=const$, $\tau=const$,
and since overall rescaling of the solution is not important
(the effective action is invariant under total rescaling,
or, equivalently, under total shift in $\tau$-coordinate),
we take it as follows:
\beq
\label{gum}
\theta=\theta_{c},\; \tau_{c}=0
\eeq
From Eqs.(\ref{bps2}),(\ref{under2}),(\ref{over2}) one immediately
obtains
\begin{eqnarray}
\label{gum2}
\cos \theta_{c}=z_{0},\nonumber\\
\sinh a = \sinh \tau_{m} \cos\theta_{c}\\
\cosh b = \cosh \rho_{m} \cos\theta_{c}.\nonumber
\end{eqnarray}

The  Eqs.(\ref{bps2}),(\ref{under2}),(\ref{over2}),
(\ref{gum}),(\ref{gum2}) specify the geometry of the
Big Bang bounce. However we still have to define charges
of the branes involved and to verify that the junction conditions are
fulfilled.

The charge conservation, with appropriate choice of orientation
of the branes, reads
\beq
\label{charge}
Q_{0}=Q_{-}-Q_{+}.
\eeq
Hereafter subscripts "0","-", and "+" indicate the  BPS brane,
the undercharged brane and the overcharged brane.
All $Q$'s are assumed to be positive.

The force balance condition obviously reads (we take projections
onto $\partial/\partial\theta$ and onto $\partial/\partial\tau$
directions):
\begin{eqnarray}
\label{balance}
T_{0} \cos\alpha_{0} + T_{-} \cos\alpha_{-}=T_{+}\cos\alpha_{+}
\nonumber\\
T_{0} \sin\alpha_{0} + T_{+} \sin\alpha_{+}=T_{-}\sin\alpha_{-}
\end{eqnarray}
where the angles are defined on fig.5. From geometry of the
picture and using Eqs.(\ref{over2}),(\ref{under2}),(\ref{gum}),
(\ref{gum2}) one obtains
\begin{eqnarray}
\label{angles}
\alpha_{0}=\theta_{c}\nonumber\\
\tan\alpha_{-}=\frac{\sin \theta_{c}\sinh \tau_{m}}{\cosh a}\\
\tan\alpha_{+}=\frac{\sin \theta_{c}\cosh \rho_{m}}{\sinh b}\nonumber
\end{eqnarray}

According to the above definition of $q$ (see Eq.(\ref{action2})) we take the
following parametrization of the tensions of the three pieces of the bounce:
\beq
\label{tension}
T_{0}=Q_{0}T,\;T_{-}=\frac{Q_{-}}{q_{-}}T,\;T_{+}=\frac{Q_{+}}{q_{+}}T
\eeq
where $q_{-}=\tanh \tau_{m}$, $q_{+}=1/\tanh\rho_{m}$.

Substituting Eq.(\ref{tension}) into the force balance condition
Eq.(\ref{balance}) and using the charge
conservation condition Eq.(\ref{charge}) one obtains a linear
system for the charges of R-branes:
\begin{eqnarray}
\label{last}
Q_{-} \left( \cos\theta_{c} + \frac{\cos\alpha_{-}}{q_{-}}\right) -
Q_{+} \left( \cos\theta_{c} + \frac{\cos\alpha_{+}}{q_{+}}\right)=0
\nonumber\\
Q_{-} \left( \sin\theta_{c} - \frac{\sin\alpha_{-}}{q_{-}}\right) -
Q_{+} \left( \sin\theta_{c} - \frac{\sin\alpha_{+}}{q_{+}}\right)=0
\end{eqnarray}
Using Eqs.(\ref{gum2},(\ref{angles}) one can straightforwardly verify 
that determinant of this linear
system is equal to zero, so the system is compatible and
defines the ratio of charges of the R-branes at which the
force balance condition Eq.(\ref{balance}) is satisfied.
This completes our construction of the Big Bang bounce.

4. We have described spontaneous creation of the brane configuration
which looks very similar to those involved in the so
called Brane World scenario.
The initial background consists of $AdS_{5}$ metric and homogeneous
4-form field (so that its curvature form is proportional to the
volume form with a constant coefficient), the final state
(in "Minkowski" picture) involves a spatially finite, flat
piece of brane at rest (RS-brane) restricted by the junction with two
expanding R-branes, the distance between RS- and R- branes and the size
of the R-branes rapidly increasing. Since RS-brane is located along the same
section of $AdS_{5}$ as the physical brane in \cite{rs2}, we take for granted
the 4d localization of gravity and the validity of 4d Newton law (modulo
decreasing finite size effects) on the RS-brane. We stress
that no negative tension branes are involved in our picture.

To conclude, we would like to point out following interesting features
of this kind of model.

i)As we already pointed out in \cite{gs}, in a model of this kind the
early Universe is five dimensional so one could expect
interesting cosmological implications.

ii)Since the RS-brane is a piece of the BPS brane, we expect
that 4d supersymmetry is broken by the finite size effects.
Moreover the scale of the SUSY breaking effects changes in time.

iii)Interesting cosmological consequences should follow the fact that
a matter, which is assumed to be localized on the branes, can penetrate
via junction manifold to/from our 4d Universe from/to the RS branes (junctions
in the Brane World context were considered in \cite{junction2}).
In particular there are models with chirality localized on the
junctions so in such case we would  have a kind of anomaly
phenomena.

iv) Finally, note that string or branes connecting RS- and R-branes would result
in the particle or extended objects with increasing masses (tensions)
in the Brane World.

The work of A.G was partially supported by grants INTAS-99-1705 and
RFBR-98-01-00327
and the work of K.S.- by grant INTAS-97-0103.

\end{document}